\documentclass{article}

\usepackage{PRIMEarxiv}

\usepackage[utf8]{inputenc} 
\usepackage[T1]{fontenc}    
\usepackage{hyperref}       
\usepackage{url}            
\usepackage{booktabs}       
\usepackage{amsfonts}       
\usepackage{nicefrac}       
\usepackage{microtype}      
\usepackage{lipsum}
\usepackage{fancyhdr}       
\usepackage{graphicx}       
\usepackage{graphicx}
\usepackage{wrapfig}

\title{Machine Learning Assisted Postural Movement Recognition using Photoplethysmography(PPG)
\thanks{\textit{This work was carried out while the first author was a student in the University of Bristol, UK.}} 
}

\author{
  Robbie Maccay and Roshan Weerasekera \\
  School of Electrical Electronic and Mechanical Engineering (EEME) \\
  University of Bristol \\
  Bristol, UK\\
  \texttt{corresponding author: roshan.weerasekera@bristol.ac.uk} \\
}

\begin{document}
\maketitle

\begin{abstract}
With the growing percentage of elderly people and care home admissions, there is an urgent need for the development of fall detection and fall prevention technologies. This work presents, for the first time, the use of machine learning techniques to recognize postural movements exclusively from Photoplethysmography (PPG) data. To achieve this goal, a device was developed for reading the PPG signal, segmenting the PPG signals into individual pulses, extracting pulse morphology and homeostatic characteristic features, and evaluating different ML algorithms. Investigations into different postural movements (stationary, sitting to standing, and lying to standing) were performed by 11 participants. The results of these investigations provided insight into the differences in homeostasis after the movements in the PPG signal. Various machine learning approaches were used for classification, and the Artificial Neural Network (ANN) was found to be the best classifier, with a testing accuracy of 85.2\% and an F1 score of 78\% from experimental results.
\end{abstract}

\keywords{Artificial Intelligence \and Fall detection \and Photoplethysmography \and Postural \and Machine Learning \and Wearables }

\section{Introduction}
People are living longer, and the aging population of the UK is ever increasing. The UK currently has a population of 5.5 million people aged over 75, which is set to increase to 7.1 million by 2035 \cite{ref1}. Of the 3.2 million of the 5.5 million people over the age of 80, half will have at least one fall a year \cite{ref2}. These falls are caused due to many factors, including muscle weakness, poor balance or visual impairment \cite{ref3}. With an elderly person falling every ten seconds in the UK, the prevalence of life altering injuries, such as head injuries and hip fractures, is high and can prove to be fatal \cite{ref4}. These injuries can lead to individuals requiring hospitalisation and surgery. They result in a loss of confidence and anxiety of falls in the future, leading to them restricting their activities in their daily lives \cite{ref4} and can result in requiring care home admission. This combination of physical and psychological impacts to the geriatric populations has led to falls being the ninth leading cause of disability-adjusted life years (DALYs) in England in 2013, putting a large strain on the National Health Service (NHS), costing £435 million annually in England alone for falls in the house \cite{ref3}.

Fall detection and fall prevention are two crucial strategies to reduce the prevalence of falls and allow the growing older population to maintain their independence. Fall detection is defined as the detection of a fall using sensors and cameras to summon help \cite{ref4}. Fall prevention refers to systems to stop falls by observing the person’s movement \cite{ref4} and actions to reduce the likelihood of falls. Fall preventative activities are carried out across many health disciplines and have been shown to have a large impact on the frequencies of falls experienced by the individuals utilising them \cite{ref5}. These methods including exercise, fall risk assessments, assistive equipment and technological based interventions \cite{ref5}. Exercise improves muscular strength and balance, and the fall risk assessments allow for clinical staff to assess mobility issues and physiological factors that may incur falls. Assistive equipment, such as grab rails and hoists, provide functional support to older adults, and aim to help with mobility around the house. Technological fall interventions are methods of identifying falls and addressing fall risks. These can be divided into pre and post fall interventions and injury prevention \cite{ref5}. Pre and post fall interventions use information sources, such as walking patterns, to recognise and alert users of fall risks. Injury prevention are systems that detect when falls occur and minimise that may occur after the event of falling. These methods reduce the frequency of falls and give people a chance to maintain their mobility. 

People in a care home are particularly at risk of falls. With the growing elderly population, the population of those that require care in nursing homes will increase. There are now over 400,00 people living in over 16,700 residential care and nursing homes in the UK \cite{ref6}. A high percentage of these have cognitive impairment that may lead to a tendency to wander and a lack of awareness of fall risks. Maintaining their safety remains a challenge, especially throughout the night when nursing staff is reduced. Therefore, effective methods for continually monitoring people remotely are a must.

Since over 65\% of falls occur in the bedroom, of which 80.1\% occur around the bed \cite{ref7}, systems that focus on people at risk of falls have been developed to allow care staff to monitor movements out of the bed. These bed exiting systems alert care staff as to if the person is attempting to leave the bed, has gotten out of bed or has fallen \cite{ref8}. The systems can involve attachments to the person (e.g. garment clips) or are part of the bed (e.g. pressure-sensitive mats placed at the side of the bed and bedside infrared beam detectors). However, the fall risks associated with standing up are similarly present if a person rises from a chair. To produce a device that could allow care home staff to monitor if a person is standing up, as sensor that could measure the postural position would be required. In this paper, a system that could follow the at-risk person around their bedroom and notify care staff if they are attempting to stand up from their stationary position is proposed. This could improve patient monitoring and allow for care staff to differentiate the current fall risk of the user.

\begin{figure}[!h]
    \centering   
    \includegraphics[width=\textwidth]{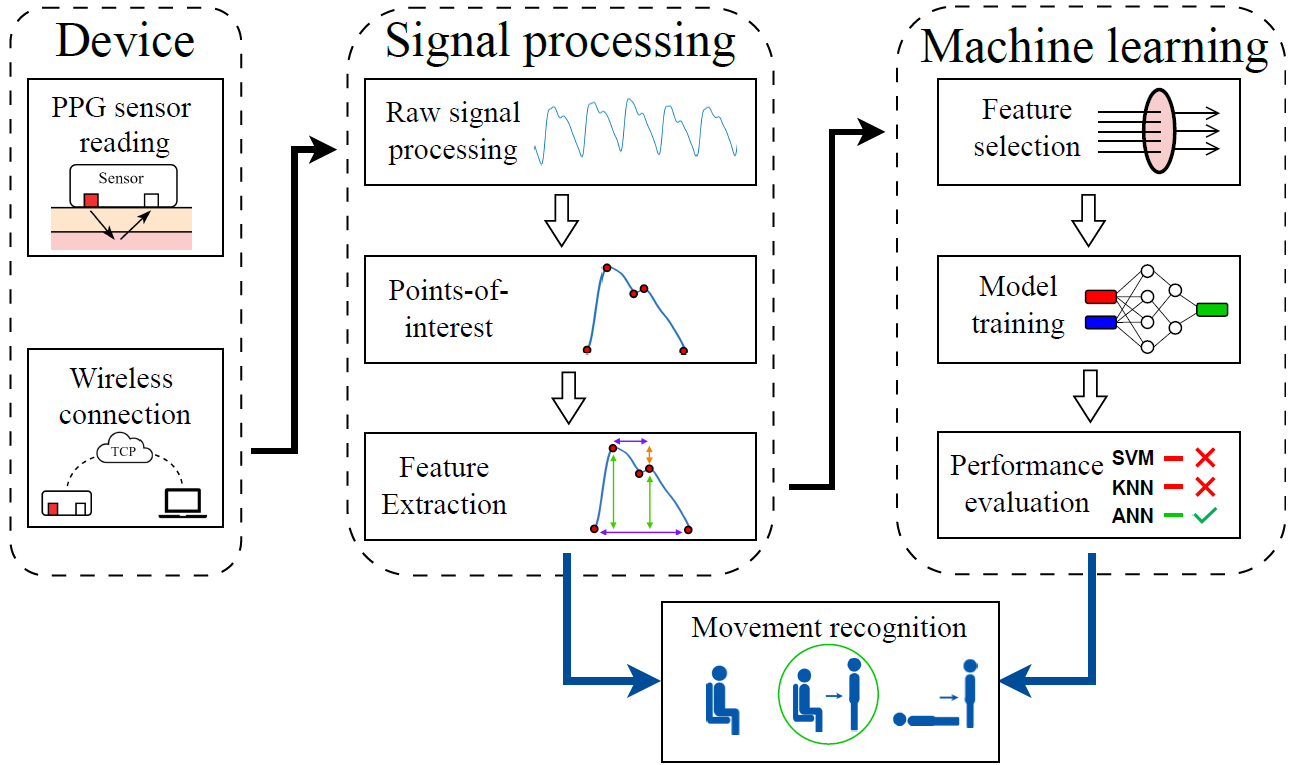}
    \caption{PPG based postural movement detection methodology}
    \label{fig:methodology}
\end{figure}

To achieve the detection of the individuals standing up, a wearable sensor detection of the movement would be required. A non-invasive, low-cost sensor that can be utilised for detecting movement is Photoplethysmography (PPG). PPG is an optical method for detecting changes in blood volume as a person moves \cite{ref9}. As the person stands up, orthostatic stress causes a sudden increase in blood volume, differing in magnitude depending on the age and health of the individual and the type of postural change \cite{ref10}. Signal analysis, segmentation, and feature extraction of PPG signals found during movement could enable a Machine Learning (ML) algorithm to classify the changes in blood volume. This process allows for the automatic recognition of postural movement. Other Human Activity Recognition (HAR) \cite{ref11,ref12,ref13} systems have used ML for recognising movement patterns in PPG signal. However, these studies have been employed for recognition of ambulatory movements, such as walking and jogging. There are no known examples in the literature of techniques proposed for postural movement recognition for a standalone PPG sensor. 

This paper is structured as follows. Section 2 give a background description on PPG, ML and HAR, as well as their current related work. Section 3 discusses the approach taken for this project, as seen in Figure \ref{fig:methodology}, for the device design, signal processing and ML. Finally, the results of the ML training and testing are presented in section 4 and section 5 is the discussion of the results and the project.

\section{Background and Related Work}
This section provides a comprehensive overview of PPG signals, highlighting their characteristics and applications in monitoring human activities and other related uses.
\subsection{Photoplethysmography}
\begin{wrapfigure}[16]{r}{0.4\textwidth}
    \centering  \includegraphics[scale=1]{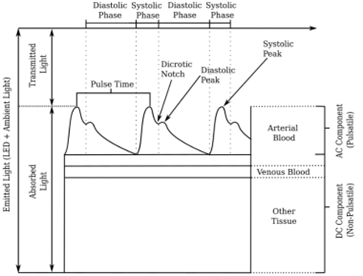}
    \caption{Principles of the PPG signal \cite{ref15}}
    \label{fig:PPGSignal}
\end{wrapfigure}
Photoplethysmography is an optical method for measuring pulsatile blood volume in the microvascular bed of tissue \cite{ref9}. The method involves the use of a light emitting source, such as an LED, to radiate light onto tissue, such as the skin. A photodetector measures the light that has been reflected from or transmitted through the tissue.  The blood volume changes are induced by the pressure pulse of each cardiac cycle \cite{ref14}, effecting the received light intensity at the photodetector \cite{ref15}. Figure \ref{fig:PPGSignal} illustrates the standard composition of a PPG waveform. The waveform can be split into two sections. These are the AC component, which represents the pulsatile component of the blood, and the DC component, which relates to the average blood volume of the tissue \cite{ref9}. The DC component changes slowly due to factors such as respiration and blood vessel vasoconstrictor waves. Each PPG pulse begins with a sharp increase in blood being pumped into the tissue. Chen et al \cite{ref16} proposed a time domain analysis method for determining the pulse onset, utilising a moving average filter to create a baseline. The onset was determined by filtering out PPG values above the baseline, removing the detection of false peaks. The Points-Of-Interest (POI) that can be found from the AC component include the systolic peak, dicrotic notch and the diastolic peak . The systolic peak marks the largest amount of blood found in the tissue. The amplitude of the pulse then decreases into a local minimum, called the dicrotic notch, indicating the closure of the aortic valves \cite{ref17}. The signal found between the onset of the pulse and the dicrotic notch is referred to as the systolic phase and the signal after the notch till the end of the wave and onset of the consecutive pulse is called the diastolic phase. An additional peak can be found after the dicrotic notch, called the diastolic peak. The shape of the waveform can differ greatly between individuals, due to factors such as age and exercise \cite{ref18}. 

\begin{wrapfigure}[14]{l}{0.45\textwidth}
    \centering
    \includegraphics{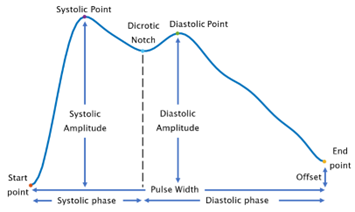}
    \caption{PPG pulse points of interest and features}
    \label{fig:PPGInterestPoints}
\end{wrapfigure}

The wavelengths of the light of the LEDs are selected depending on the purpose for the sensor and the placement of the sensor on the body. The wavelengths of the light of the LEDs are selected depending on the types of measurements that are completed by the device. The largest components of the upper layers of tissue that affect the rates of absorption of light are water, haemoglobin and tissue depth \cite{ref9}, \cite{ref15}. Water is a major component of the skin and is highly absorbent to ultraviolet and higher frequency light signals, whereas lower frequency and Infra-Red (IR) signals are in a window of the absorption spectra, allowing for higher penetration of light into the microvascular bed of tissue. Oxygenated and deoxygenated Haemoglobin have significant different absorption rates, expect at wavelengths close to infra-red, meaning that green LEDs are commonly used for oxygen measurements \cite{ref19}. To maximise light penetration and PPG received intensities, green and blue LEDs should be used for wrist applications, whereas lower frequency light, such as red, should be used for forehead applications \cite{ref20}. Skin temperature also has a quantitative effect on the signal quality, as lower temperatures reduces perfusion rates in the vascular bed, as the body attempts to conserve body heat, constricting blood vessels and increasing the distance the light has to travel \cite{ref15}. The PPG sensor contact pressure (CP), which is the external pressure applied by the sensor on the surface of the skin, further impacts the PPG readings \cite{ref21}. The amplitude and shape of the PPG waveform are impacted, with higher contact pressures increasing the accuracy of physiological measurements, such as the heart rate\cite{ref18}.Transmission mode detectors are commonly used for PPG sensors placed at the ear or finger. Reflection mode sensor allow for a higher level of flexibility, such that they are more suitable for wearable devices, and are used in ankle, chest, forearm, forehead and wrist applications \cite{ref19}. Research has also been done for multi-wavelength PPG sensors \cite{ref15}, where using combinations of IR, red and green LEDs are used in close proximity to each other and around 2mm from the photodiode. 

The site of the sensor on the body is an important consideration for the mode, wavelength and sensor used and can lead to different morphology of the waveform. The sites that are primarily used are the ankle, chest, earlobe, forearm, forehead, finger and upper wrist. A large number of the consumer devices use wrist applications \cite{ref18}, \cite{ref19}, as it is convenient to wear, highly portable and can be combined in smart watches. The upper wrist has been found to have relatively low signal amplitude in comparison to other sites \cite{ref19} and a higher susceptibility to Motion Artifacts (MA). Another key placement of the PPG sensor is the ear lobe, as the lack of cartilage and thus contain a large volume of blood, creating a suitable site for transmission type sensors. However, their usage can become uncomfortable after long monitoring periods \cite{ref22}. The finger is a common location for pulse oximetry \cite{ref23} as either a clip or ring. The forehead is suited for reflective mode sensor, as the thin skin, high density of blood vessels and skull give a powerful optical signal. The forearm can be used for arterial PPG measurements, but has been found to have a lower pulse amplitude than wrist or finger sensors \cite{ref22}. From the PPG waveform, pulse time domain analysis can be completed to extract features of the signal \cite{ref9}, \cite{ref24}. From the POI from each of the pulses of the wave, as discussed earlier, extracted features, such as the pulse width, systolic and diastolic amplitudes, pulse offset, and peak difference can be found. Figure 3 illustrates the correlation between the extracted features and the POI.

The pulse width is described as the time difference between the onset of two consecutive pulses. The systolic amplitude is defined as the difference between the height of the systolic peak and the pulse onset, while the diastolic amplitude is the difference in height between the diastolic peak and the start point. The difference in magnitude between the pulse onset and end point is defined as the pulse offset. The peak difference is the difference in systolic and diastolic magnitudes.  

Further features of the PPG waveform can be extracted from the first and second derivatives of the waveform \cite{ref18}, \cite{ref19}, \cite{ref25}. The first derivative can be used for reliable detection of the diastolic peak, while the second derivative can be used for detection of the dicrotic notch \cite{ref17}. The features of the second derivative can be used for diagnosis of cardiac abnormalities \cite{ref19}, such as arterial stiffness and arterial aging. Further features can be found when combining the PPG sensor with an ECG device, such as the Pulse-Transit-Time (PTT) and the Pulse-Wave-Velocity (PWV). PTT is defined as the time between the R-peak of the ECG to the onset of the PPG wave. The PPT can also be derived from PPG signals from two sensor placed at different sites on the body \cite{ref17}. The PWV is defined as the inverse of the time between the R-peak in the ECG to the peak in the first derivative of the PPG signal. Furthermore, frequency domain analysis can be utilised for stationary processes, but should not be used for monitoring the morphology of the pulse due to movement \cite{ref10}.

\begin{wrapfigure}[18]{l}{0.4\textwidth}
    \centering
    \includegraphics[scale=1]{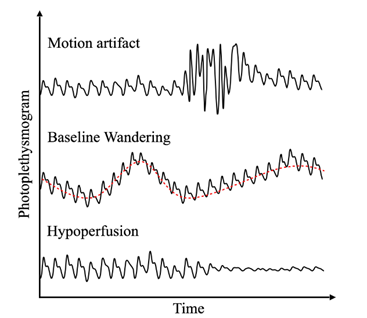}
    \caption{Examples of distortions in the PPG signal due to motion artifacts, baseline wandering and hypoperfusion \cite{ref15}}
    \label{fig:PPGdistortions}
\end{wrapfigure}

To give confidence in the accuracy of the physiological measurements made by a PPG device, reproducibility is an important metric in the development of the sensor. For continuous monitoring, PPG waveform noise can reduce the accuracy and reproducibility of the measurement system. The characteristics of the noise that can affect the PPG analysis are Motion Artifacts (MA), baseline wandering and hypoperfusion \cite{ref9}, \cite{ref17}. Figure \ref{fig:PPGdistortions} illustrates the three PPG distortions.

As changes in the body movements occur, such as hand movements or walking, the volume of blood adjusts. These sudden changes in the blood can create distortions in the PPG waveform, called Motion artifacts \cite{ref17}. If the body part in which PPG is placed on moves during measurements, the relative movement between the PPG LED and photoreceptor, as well as the movement of the skin in contact with the sensor, can also cause distortions in the waveform \cite{ref11}. Baseline wandering is noise due to changes in the baseline of the pulsatile component and the AC amplitude of the PPG signal \cite{ref17}. Respiration, sympathetic nervous system activities and thermoregulation are all factors that can move the baseline \cite{ref17}, interfering with the analysis of the AC component of the PPG. Hypoperfusion is a reduction in the changes of blood volume in the blood vessels due to factors such as hypothermia and vasoconstriction \cite{ref17}. 

From the features extracted from the waveform, PPG allows for the detection of cardiovascular and respiratory physiological parameters. Blood Oxygen Saturation; Heart Rate; Pulse Rate Variability; Blood pressure; Respiration Rate; Arterial stiffness. 

Using these physiological parameters, doctors can diagnose a wide range of clinical conditions. Benefits of the use of PPG lies in the non-invasive measurements and the convenience of wearable devices for the assessment of vascular and autonomic function \cite{ref9}, \cite{ref18}. One of the early uses for PPG was to determine the SPO2 for individuals under anaesthesia \cite{ref23}. The derivatives of the PPG signal can be used to determine the cardiovascular mortality \cite{ref23}, allowing for the identification of the individuals at risk of cardiovascular diseases. PPG devices are used as part of an overnight test to help diagnose sleep disorders, including Obstructive Sleep Apnoea (OSA) \cite{ref23}. By observing changes in the SPO2 and changes of the PPG signal with respiration are often utilised for diagnosis \cite{ref18}. Other conditions include detection of atrial fibrillation, monitoring the spread of infectious diseases, blood sugar monitoring for diabetics and neurological diagnosis and treatment \cite{ref9}, \cite{ref18}, \cite{ref23}. 

When a person stands up, changes in the blood flow in the microvascular blood vessels occurs. When a person rises, orthostatic stress occurs, where blood is forced to the lower extremities and pools in the peripheral veins \cite{ref10}.This change in blood volume causes a drop in blood pressure and reduced cardiac output. When this occurs, baroreceptors in the arteries will detect the change in blood pressure and will initiate a baroreflex response. This response will increase heart rate and constrict the venous system to the peripheral arterioles. This reflex restores the blood pressure back to a nominal value. Conditions, such as Orthostatic Hypotension (OH), defined as a BP drop of at least 20 mm Hg systolic and/or >10 mm Hg diastolic 3 minutes after postural change \cite{ref26}, can effect the magnitude of the blood pressure drop and the time it takes for the blood pressure to return to nominal values \cite{ref27}. OH is prevalent in older adults and has a high associating with falls \cite{ref26}. 

\subsection{Machine Learning}
Machine learning (ML) is a tool that is used widely for fall identification and prevention, where activities such as falling forwards, falling backwards and spinning can be differentiated from non-fall activities \cite{ref4}. ML is a learning ability for a system to observe trends in data. The steps required for the development of ML involves data collection, data processing, feature extraction, training the ML algorithms and performance evaluation. For PPG, the raw data gathered from the sensor is noisy and processing techniques can be used to clean the signal. Feature selection is an important step that can influence the classification accuracy of the models and can reduce the processing costs in model training [4]. A large number of features can lead to overfitting of the model \cite{ref4}, reducing the effectiveness of the model in predicting unseen data and reducing the testing accuracy performance. Therefore, assessment of the importance of predictors and the removal of redundant features is an essential step. Machine learning can be categorised as non-deep learning or deep learning methods \cite{ref11}. Non-deep learning approaches include Random Forest (RF), decision tree, Support Vector Machines (SVM) and k-Nearest Neighbour (KNN). Deep learning approaches include Deep Neural Networks (DNN), Convolutional Neural Networks (CNN) and Recurrent Neural Networks (RNN). 

This can be done Following feature extraction from PPG signals, the data is partitioned into training and testing sets, with the ratio determined by system design considerations. The training data is utilised by the ML algorithm to discern data patterns, such as fall activities for fall prevention recognition techniques. The testing data is used to evaluate the performance of the algorithm, using accuracy, sensitivity and specificity metrics. 

ML has found a wide array of applications in the analysis and interpretation of PPG signals. It facilitates the automatic detection of physiological parameters including HR \cite{ref28}, BP \cite{ref29}, \cite{ref30} and blood glucose \cite{ref30}.Goh et al described a deep learning approach for the automatic detection of MA in PPG signals \cite{ref31}, allowing for the automatic cleaning of PPG signals for physiological measurements. Alkhodari et al utilised ML for the recognition of cardiovascular risk factors \cite{ref32}.
The ML algorithms for PPG analysis in this project are as follows:
\paragraph{Discriminant Analysis (DA):}
Discriminant analysis is a technique used mainly in the transformation of high-dimensional data into a lower dimensional format[4], thereby maximising the class separability \cite{ref33}. Discriminant analysis is a good model for prediction of wide datasets. The most popular approach is for Linear Discriminant Analysis (LDA) \cite{ref4}, where linear boundaries between classes are found, while quadratic discriminant analysis finds non-linear separations. 

\paragraph{Naive Bayes (NB):}
The Naïve Bayes utilises the Bayes Theorem to construct classes based on probability \cite{ref4}. The algorithm determines the probability of each class in the training data and the likelihood of observing features from a test dataset for a given class. The Bayes theorem combines these values to determines the probability of the different possible classes for the test dataset. The class of the highest probability is chosen as the predicted class. This is a popular method used to provide fast predictive results.
\paragraph{Support Vector Machines (SVM):}
Support Vector Machines (SVM) is a supervised machine learning model that attempts to find a hyperplane that can divide classes of data, such that there is an equal distance between classes to the hyperplane \cite{ref4}. The algorithm attempts to find the optimal hyperplane to maximise the margin between data, so that the hyperplane acts as a decision boundary between classes [4]. The shapes of the hyperplanes can either be linear or non-linear. The linear classifier finds boundaries between linearly isolated data points\cite{ref4} and the non-linear classifier finds hyperplanes between more complex class boundaries. The performance of SVM depends on the magnitude of the separation between classes \cite{ref34}. Therefore, in cases where class boundaries overlap, the accuracy of the model reduces.

\paragraph{Decision tree (DT):}
Decision tree (DT) models are structured as a series of decision trees \cite{ref30}. Each decision tree consists of nodes, representing a test, with branches sprouting from each node, representing the outcomes of the test. The classifier is built based on the path that the algorithm takes through the tree.  DTs are very sensitive to training data, where small changes in the training data can affect the overall tree structure \cite{ref30}.

\paragraph{K-Nearest Neighbours (KNN):}
K-Nearest Neighbours (KNN) is a supervised learning algorithm that can be used for both classification and regression models \cite{ref4}. This algorithm assumes that points of the same class are found in close proximity to each other. The first step of the algorithm is to determine the distance between the training data to the new data point that the algorithm is attempting to classify. The test points are then ranked based on their proximity to the new data point. The first k number of neighbours closest to the new data point are found and the algorithm determines the average class of these neighbours. The algorithm then assigns the most common class identifier to the new data point.

\paragraph{Artificial Neural Network (ANN):}
Artificial Neural Networks (ANN) is a ML algorithm inspired by neurology in the human brain \cite{ref4}. ANNs are made up of processing elements, referred to as neurons, which are connected together and are organised into layers \cite{ref35}. Each of these connections has a weight, determining the strength of the connection. ANN utilises an activation function to translate a set of inputs, known as the input layer, to a set of outputs, known as the output layer. The activation function takes the sum of each input after multiplying them with their respective weight. Common activation functions include sigmoid, Rectified Linear Unit (ReLU), and the hyperbolic tangent (tanh). ANN can also have hidden layers between the input and output layers to cope with nonlinearly separable problem \cite{ref35}. For hidden layers, the output of a ANN signal can be used as an input for a consecutive ANN.

The training of the ANN comes in the adjustments of the weights. For each input, a trainer is used to predict the input and create a cost function for each layer, finding the difference between the predicted and actual values. ANN repeats the training process and adjusts the weights to reduce the difference between the predicted and actual values. The result is a system that can take input data and create a prediction, based on the patterns learned during training. 

\subsection{Human Activity Recognition}
Smart watches, phone movement and other wearable sensors are used for recognition of various human activities, such as jogging and running \cite{ref11}. Human Activity Recognition (HAR) is used in health care, skills assessments and surveillance \cite{ref12}. There are a number of types of sensors used for HAR, chosen depending on factors such as the targeted activities, costs and form factor \cite{ref11}. The most popular sensor is the tri-axial accelerometers, which can be found in many smartphones and smart watches \cite{ref4}, \cite{ref11}. Chou et al \cite{ref36} proposed a bed exiting alarm system that utilised a singular accelerometer placed on the chest that monitored the tilt angle of the upper body. For posture related tasks, waist and thigh accelerometer combinations allow for the greatest accuracy \cite{ref37}. Accelerometers, as well as other types of inertial sensors, can also be used in conjunction with PPG sensor for effective tracking of movement actions \cite{ref13} and as a reference signal, utilised in the cancellation of MA in the PPG signal \cite{ref24}. The advantage of using PPG over accelerometers is the lower power consumption \cite{ref11} and the ability to computing HAR while measuring other physiological parameters, such as the heart rate \cite{ref11}, \cite{ref12}. This allows for the reduction in resources while sensing. 

ML have proven to be a successful approach for HAR. For fall prevention, studies involving SVM have been the most popular \cite{ref4}. This is due to SVMs ability to detect distinct datasets, with the differences between the data of a standing and fallen person creating a wide margin around the hyperplane. For PPG signal HAR using ML, Mehrang et al \cite{ref34} used Randon Forest (RF) and SVM algorithms for a PPG and accelerometer combination study to detect sitting, household activity and low and high intensity cycling. This approach resulted in 89.2\% and 85.6\% average recognition accuracies for the RF and SVM classifiers, respectively. Various deep learning approaches have been utilised for HAR using PPG data. Alessandrini et al \cite{ref12} utilised a RNN for the detection of resting, squatting and stepper activities utilising a PPG and accelerometer combination data set. Hnoohom et al \cite{ref38} produced a deep learning approach with an F1 score of 90\% for various exercise and movement activities. For standalone PPG recordings, CNN \cite{ref31} and DNN \cite{ref11} approaches for various ambulatory activities have also achieved high levels of accuracy.

\section{Method ana Analysis}
The methodology adopted  in this work is illustrated in Figure \ref{fig:methodology}. It encompasses three primary stages: hardware/software development, signal processing, and machine learning. A detailed discussion of each stage is provided in this section. 
\subsection{Hardware Design}
\begin{wrapfigure}[14]{r}{0.33\textwidth}
    \centering
    \includegraphics[scale=0.9]{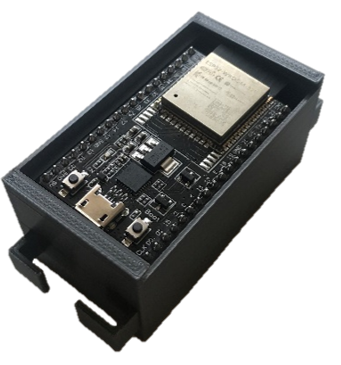}
    \caption{PPG Datacollection Wearable}
    \label{fig:PPGWearable}
\end{wrapfigure}


To monitor the PPG signal from an individual during movement, the device depicted in Figure \ref{fig:PPGWearable} was developed. The design emphasized simplicity, reliable data readings, wearability, and wireless capabilities. To ensure flexibility in modifying the design and to streamline the development process, the electronics are modular, and development kits were used where possible to demonstrate the concept. The MAX30101 \cite{ref39} , an integrated pulse oximetry and heart-rate monitor module, was chosen as the sensor. The microcontroller for this design was selected to support peripheral interfaces for sensor interaction, have low power consumption for use with a small battery, and provide wireless connectivity. The ESP32-DevKitC V4 (ESP32) \cite{ref41} was selected for this purpose. It uses the ESP32-WROOM-32 MCU module to control the sensor's operation and includes the I²C peripheral for interfacing with the MAX30101. Both the sensor and the microcontroller is powered by a battery and the entire system is stored in a 3D printed housing. Integrating the sensors and the processor into a single chip could reduce the size of wearable by approximately 50\%. However, the primary focus of this study is to explore feasibility, so the size of the wearable was not a major consideration.

To gather PPG sensory data and to transmit the data to a computer, firmware for the ESP32 and a software program on the computer were developed. The firmware involved establishing a connection to the MAX30101 and a wireless connection to the computer. The software on the computer received the data from the device and securely stored it onto the device. A TCP connection was chosen, as it allowed for remote connectivity to the device and reliable delivery of the sensory and time data. The TCP client and TCP server were found on a computer and the device respectively.

\subsection{Software Design}
\begin{figure}[!h]
    \centering
    \includegraphics[width=\linewidth]{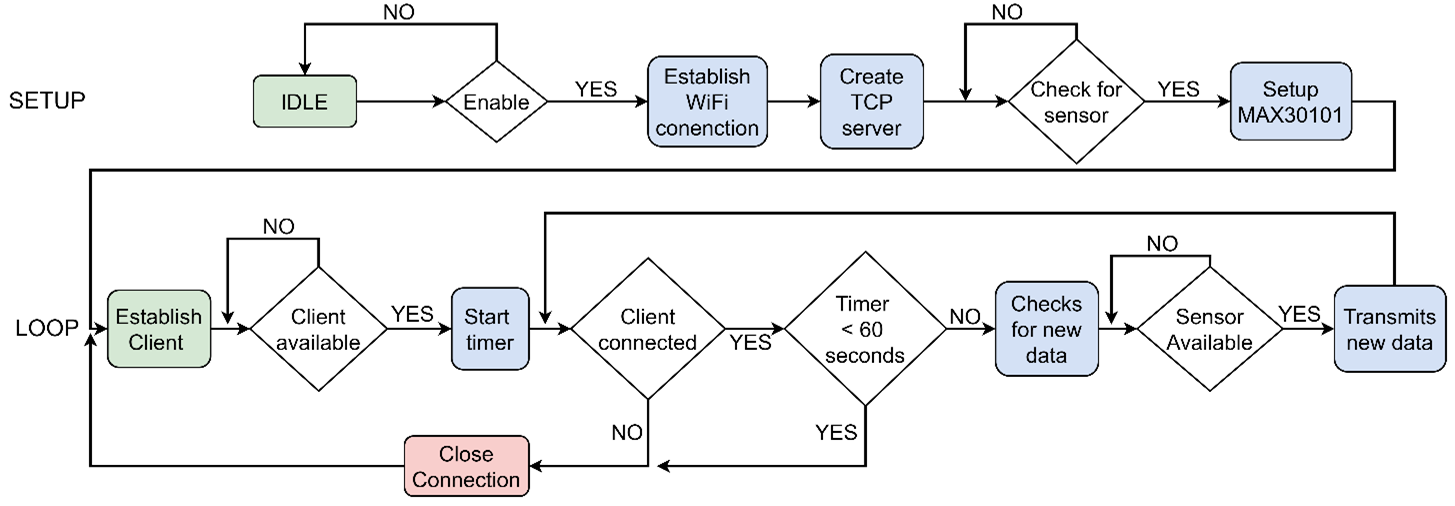}
    \caption{Flow chart for the device software (\textit{device.ino})}
    \label{fig:flowchart1}
\end{figure}

The software for the ESP32, programmed in C, was designed for two tasks: sensory data acquisition and TCP communication. The former involved establishing the I²C connection between the MAX30101 sensor and the ESP32 and reading PPG data from the red LED. The latter established the TCP server on the device and transmitted the PPG data and the time for each recording over the connection. The Arduino Wi-Fi \cite{ref42} and Wire \cite{ref43} and the SparkFun MAX30101 \cite{ref44} sensor libraries were used for this code Figure \ref{fig:flowchart1} illustrates the firmware on the ESP32.
To activate the setup of the device, the device needed to be powered and the enable button on the ESP32 had to be pressed. The Wi-Fi connection to device was then established, thus allowing for the creation of the TCP server on the device. The program then checked if there was an I²C connection between the pins of the MAX30101 and the ESP32. Upon confirmation of a connection, the parameters for the MAX30101 were set.
Upon completion of the setup, the loop code was initialised. The establishment of the TCP client object in the loop allowed for the device to check for client connections to the server. The program would then go into a nested while loop, where it continually checks that the connection has been maintained and that less than a minute of recording has taken place, otherwise closing the connection. If these conditions have been met, then the ESP32 will ask the MAX30101 to produce new data and will transmit the PPG sensory data and the time data, once they have become available. For the client to be able to differentiate the data type being transmitted, the Most Significant Bit (MSB) of the data point is set as either high for time data, or low for sensory data. The sensor and time data are transmitted as a continuous stream of 32-bit integers to the client and are separated through the use of a comma. 

\begin{figure}[t]
    \centering
    \includegraphics[width=\linewidth]{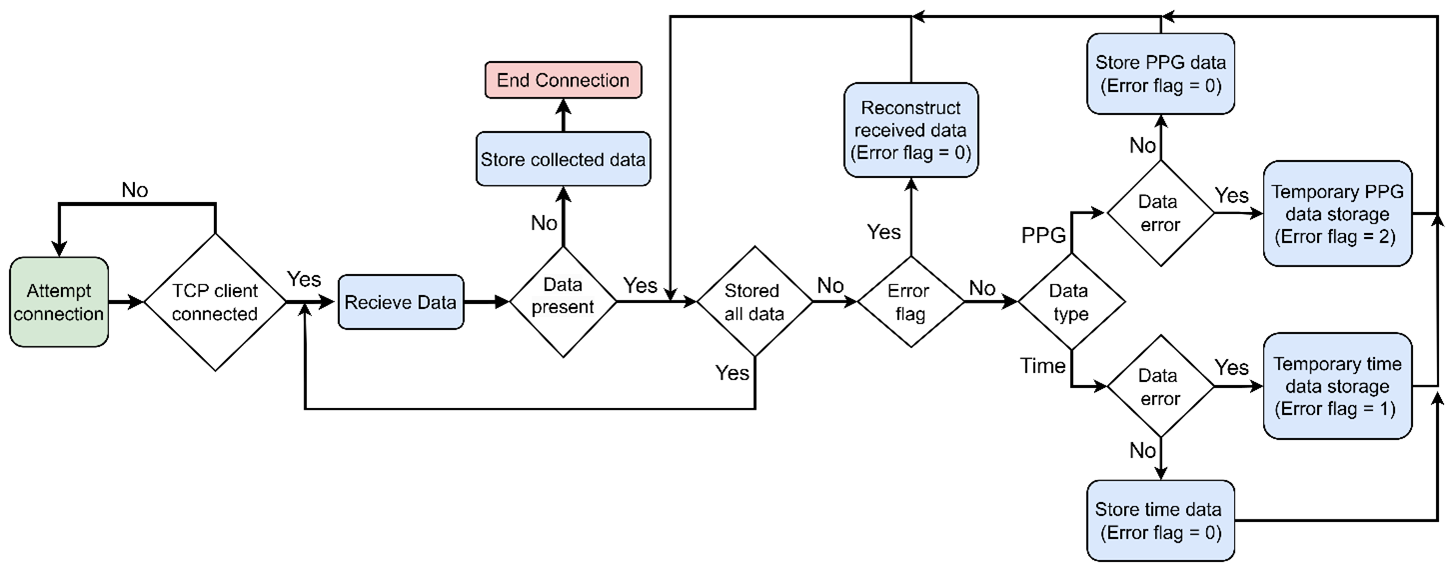}
    \caption{Flow chart for the receiver software (\textit{recciever.py})}
    \label{fig:flowchartRX}
\end{figure}

The client side of the TCP connection is a Python program that runs on a computer and utilises Socket python library \cite{ref45}. Figure \ref{fig:flowchartRX} shows the structure of the code.  The initial step created a “socket” TCP client object and attempted to connect this to the ESP32 TCP server. This will then begin a while loop that will receive data from the device until the timer on the ESP32 reaches 1 minute and closes the connection. Once connected, the client will read the data received from the device and decodes the packets into individual data points. The processing involves removing the MSB of the received data, indicating the data type, and storing the remaining data. The error flagging allows for the realignment of erroneous received data that were fragmented during the read process. If no information is being transmitted from the device, primarily due to the recording being completed, then the computer will disconnect the client to the server and will store the gathered data as a .m file.

\begin{figure}[!h]
    \centering
    \includegraphics[width=0.8\linewidth]{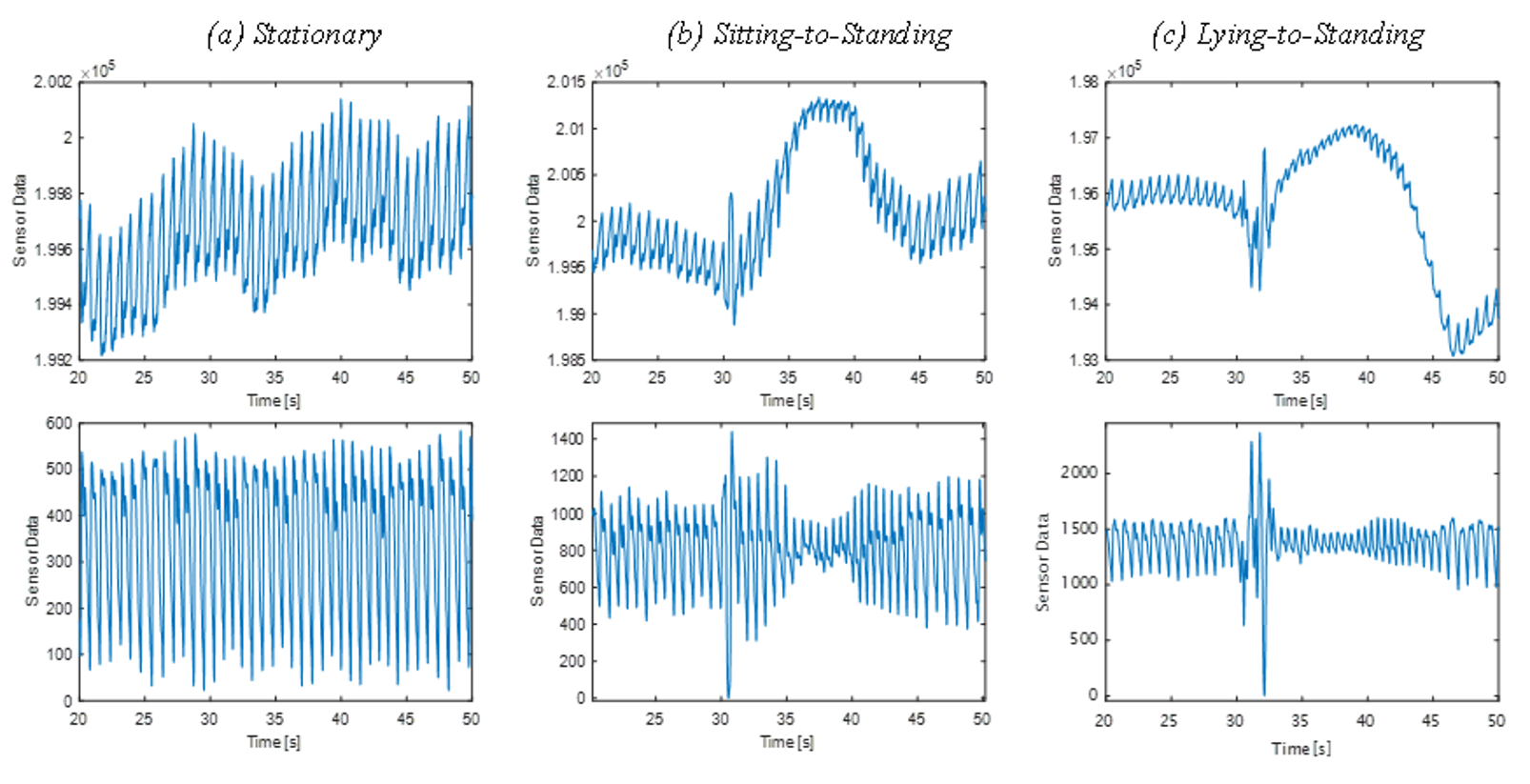}
    \caption{Examples of the raw and filtered PPG signals for the postural movements}
    \label{fig:17}
\end{figure}

\subsection{Data Collection}
Participants were recruited to take part in a controlled postural movement recognition experiment. Participants were asked to perform two activities. These movements included: 
\begin{enumerate}
    \item Sitting stationary on a chair to move into a standing position.
    \item Lying flat on the ground to move into a standing position.
\end{enumerate}
The initial step was for the participant to place the device onto their temple (other locations). Adjustments were made to the position of the device to ensure a good contact pressure between the device and the skin. An initial test recording was done to make sure that an expected PPG value was being read, confirming the correct placement of the device. The participant was then asked to remain in the initial stationary position for a period of 30 seconds, maintaining a regular breathing pattern and without moving or talking. At 29 seconds, an indication that the 30 second point was approaching was giving and at 30 seconds, the participant was asked to stand. They remained for a period of 30 seconds in the standing position.

11 participants were recruited for the study. The device was used to record the PPG signal whilst participants moved from sitting-to-standing and lying-to-standing. This generated a total of 2460 seconds of labelled data. Of the 41 recordings, 24 were for sitting-to-standing transitions and the remaining were for lying-to-standing transitions. Improper placement of the sensor, movement of the sensor as the participant moved into the standing position and movement during the stationary phases of the recordings created errors which meant that 3 of the recordings were not used in the data. 

Figure \ref{fig:17} illustrates the differences that can be found between the three classes. Between the postural movements changes to the signal due to the movement can be observed. For the sitting-to-standing and lying-to-standing movements, an increase in blood volume after the movement at 30 seconds and subsequently dropped at 40 seconds. Between the three movements shown in Figure \ref{fig:17}, the standard deviations of the systolic peaks of the stationary, sitting-to-standing and lying-to-standing are $\pm 160.6$, $\pm 429.7$ and $\pm 1100.7$ respectively. The greater variability of the raw signal with movements indicates that there is an increasing variability in the blood volume changes, with the largest changes occurring with lying-to-standing movements. 

\begin{figure}[!h]
    \centering
    \includegraphics[width=0.8\textwidth]{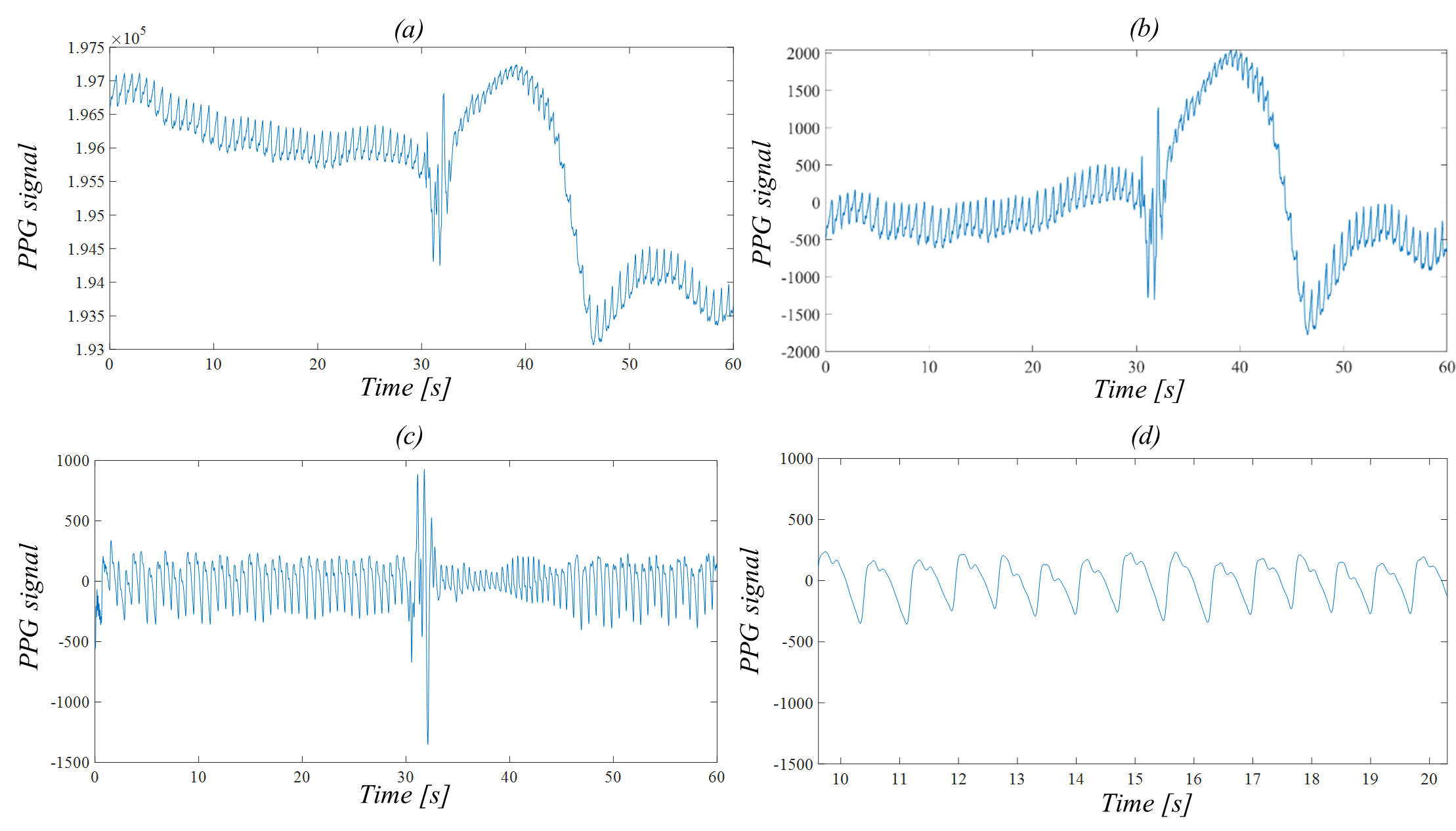}
    \caption{PPG signal processing a lying position to a standing position (a) Raw signal, (b) Detrended signal, (c) Filtered signal, (d) Filtered PPG pulses.}
    \label{fig:12}
\end{figure}

For the filtered waveforms, the changes from the stationary signal pattern were primarily during the initial movement. Additionally, the increase in blood volume is indicated by a decrease in the systolic and diastolic amplitudes, fluctuating depending on the gradient of the blood volume changes. Further differentiation between the sitting-to-standing and lying-to-standing movements can be seen in the larger amplitudes of the initial MA for the lying-to-standing movement.

\subsection{Signal Processing}

\begin{wrapfigure}[25]{R}{8cm}
    \centering
    \includegraphics[width=8cm]{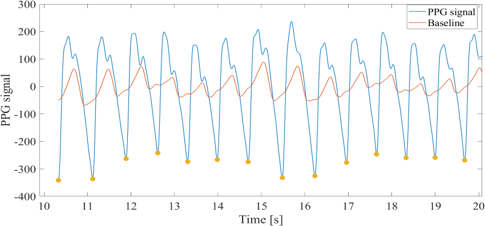}
    \caption{Example of pulse onset detection of the PPG signal}
    \label{fig:13}

    \includegraphics[width=8.3cm]{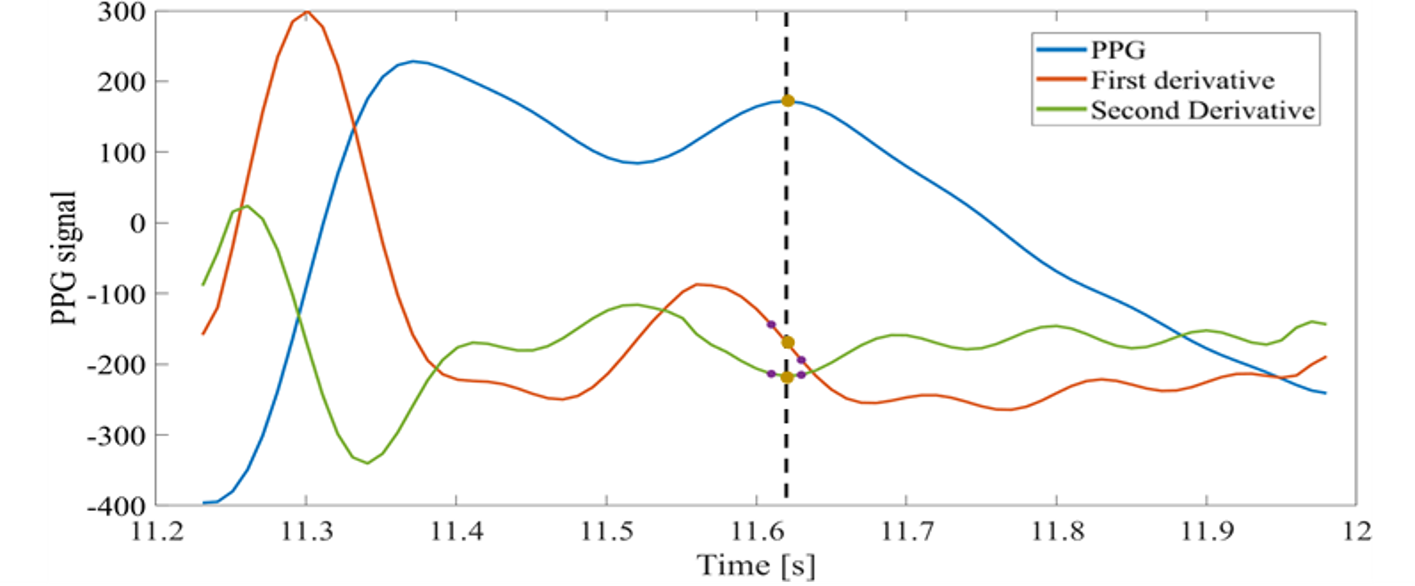}
    \caption{Diastolic peak detection, displaying the matching of the second derivative minima to the zero crossing points in the first derivative}
    \label{fig:Fig14}
    
\end{wrapfigure}

For each recording of a participant’s postural movements, the PPG information was obtained by an algorithm developed in MATLAB. The program processed the raw data into different formats, filtered unwanted noise and segmented the signal into individual PPG pulses. Each pulse then had the POI and features determined. 

To ensure that accurate results could be collected from the PPG wave, the raw signal was pre-processed before feature extraction. To remove erroneous data that may be present in the time and sensor data, results out with the expected ranges from the data were removed. Figure \ref{fig:12} shows an example of a transition from a lying position to a standing position. To isolate the MA aspects of the wave, the MATLAB ‘detrend’ function and a bandpass filter is applied to the PPG wave. The detrend function removes the linear trends and DC component of the signal. For the filter, a normalised start frequency 0.0075 Hz/sample and a stop frequency of 0.2 Hz/sample. The removal of the higher frequency aspects of the PPG waveform reduce noise and the likelihood of false peaks. The removal of the low frequency components removes the effects of baseline wandering. The affect of the filter can be seen in Figure \ref{fig:12}(c), where the PPG signal magnitudes are normalised around zero.

For the analysis of the waveform, the PPG signal was segmented into individual pulses. This was completed by identifying the pulse onsets and storing the waveforms found between these values. To identify the pulse onset, the minima of the PPG wave were identified. To improve the reliability and precision of the identification of the true pulse onsets, the process illustrated in Figure \ref{fig:13} was used. A centre moving average filter was used to calculate a baseline of the PPG waveform. To remove falls onsets, the minima of the wave were selected if the distance between minima was larger than half the period of the signal pulses and if the minima could be found below the baseline. This eliminated dicrotic notches being misidentified as the onset of the pulses.

For each of the pulses of the PPG signal, the POI were identified. For the filtered PPG signal, the POI found seen in Figure  \ref{fig:PPGInterestPoints}  were found. This included the magnitudes and times for the onset of the pulse, the end point, the systolic point, the diastolic point and the dicrotic notch.  These points were also identified for each of the pulses in the raw and detrended PPG signals. The onset and ending points were determined using the minima located during the segmentation process. The systolic point was identified by finding the maximum peak of the pulse. Figure \ref{fig:Fig14} shows the detection process for the diastolic peak.

The initial step was to determine the minima of the second derivative and the zero crossing points of the first derivative. To determine the diastolic peak, where a minima of the second derivative had the same time value as the right most zero crossing point, excluding the ending of the pulse. If a match cannot be found, the location of the right most zero crossing point is used. This increased the reliability of detecting the diastolic point, as the shallow gradient of the pulse can peak identification errors when utilising regular peak detection.

\subsection{Feature Selection and Extraction}
\begin{wrapfigure}[13]{r}{8cm}
    \centering
    \includegraphics[width=8cm]{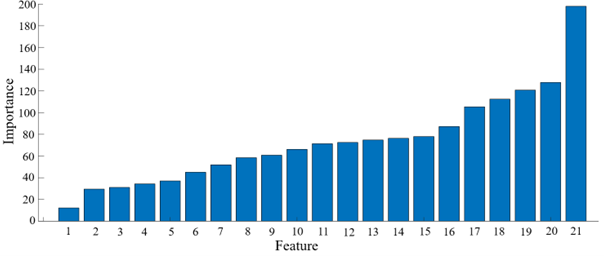}
    \caption{Chi-squared test feature selection importances. The features can be found in Table \ref{tab:Features}.}
    \label{Fig:18}
\end{wrapfigure}

To visualize the importance of the features for the removal of redundant predictors, the Chi-square test algorithm was utilized \cite{ref38}. The Chi-squared test performs a statistical hypothesis test and determines the differences between the observed and expected frequencies of samples within the data. The relevance’s of the features were then ranked. Figure \ref{Fig:18} shows the chi-squared test importance values for each of the 21 features gathered. The features that had the highest associations to the target variables were the systolic magnitude, systolic rise gradient and raw systolic amplitude. The features that had the least significance for the classifier were the detrended diastolic amplitude, diastolic magnitude and the dicrotic magnitude. The low number of features meant that a reduction of a single feature could affect the overall performance of the algorithm training, so only features that had little importance were removed. Therefore, the dicrotic magnitude was removed.

\begin{table}
\centering
\caption{Chi-squared test features in order of importance} \label{tab:Features}
\begin{tabular}{l|l}
\hline 
Feature index	&	Feature	\\ \hline
1	&	Systolic magnitude	\\ \hline
2	&	Systolic rise gradient	\\ \hline
3	&	Raw systolic amplitude	\\ \hline
4	&	systolic amplitude	\\ \hline
5	&	Systolic and diastolic peak difference	\\ \hline
6	&	Detrended systolic amplitude	\\ \hline
7	&	Pulse onset magnitude	\\ \hline
8	&	Raw offset	\\ \hline
9	&	Raw orthostatic magnitude	\\ \hline
10	&	Pulse Width	\\ \hline
11	&	End point magnitude	\\ \hline
12	&	Detrended offset	\\ \hline
13	&	Diastolic amplitude	\\ \hline
14	&	Raw diastolic amplitude	\\ \hline
15	&	Systolic phase	\\ \hline
16	&	Diastolic phase	\\ \hline
17	&	Offset	\\ \hline
18	&	Detrended orthostatic magnitude	\\ \hline
19	&	detrended diastolic amplitude	\\ \hline
20	&	diastolic magnitude	\\ \hline
21	&	dicrotic magnitude	\\ \hline

\end{tabular}
\end{table}
 
For the detection of the differences in PPG signals from individual segments, a set of features capable of discriminating between movement types is essential. These features were extracted from the POI determined for both the raw, detrended and filtered pulses of the PPG signal. The analysis done was completed using time domain analysis, as this allowed for the observation of changes in the morphology in the PPG pulses. 

For each of the pulses, the POI can be used to derive a variety of features. The features found in Figure  \ref{fig:PPGInterestPoints}, such as the pulse width, the systolic and diastolic amplitudes, the systolic and diastolic phases and the offset. Furthermore, the peak difference between the systolic and diastolic points, referred to as the PPG peak difference, and the gradient of the systolic rise were determined. Similar features were extracted from the detrended and raw PPG signals.

\begin{wrapfigure}[15]{r}{8cm}
    \centering
    \includegraphics[width=8cm]{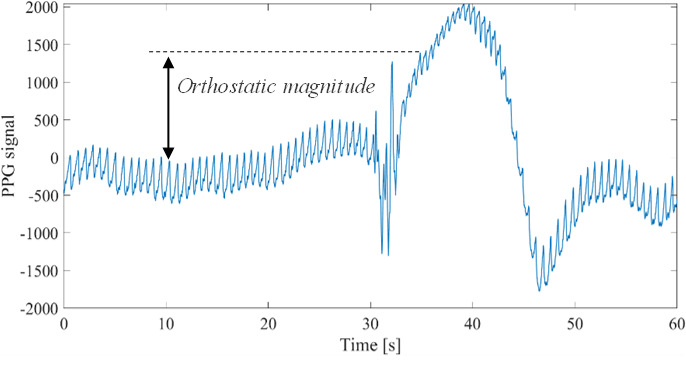}
    \caption{Example of the orthostatic magnitude for a lying-to-standing PPG signal.}
    \label{fig:15}
\end{wrapfigure}

Additionally, another feature was extracted to measure the effect of the orthostatic stress. As seen in Figure \ref{fig:15}, large changes in the systolic magnitudes of the raw signal between the PPG values at rest compared to the values after the movement. To quantify this, a feature called the orthostatic magnitude was found. A baseline value from when the wearer was stationary in either the sitting or lying positions was determined. The orthostatic magnitude is defined as the difference in height between the systolic peak and the baseline value. The orthostatic magnitude was found for both the raw and detrended signals.

The MATLAB classification learner app was utilised to train and analyse a variety of ML algorithms. A table containing the extracted features from the PPG data, where the type of movement is labelled, is input into the toolbox. The algorithms were implemented using several preset hyperparameter configurations. The data set was split a training set and a test set into a 9:1 ratio. A cross-validation strategy was applied to create divisions in the data, where k folds in the dataset are trained in the model k times, reducing the effects of overfitting. The cross-validation process split the dataset into 10 separate folds.

\subsection{Machine Learning results}
The performance of the ML models were evaluated with three metrics, accuracy, precision, sensitivity and the F1 score for each activity\cite{ref38}. The definitions for the above metrics are defined as below:
\begin{equation}
    Accuracy (\%) = \frac{TP+TN}{TP+FP+FN+TN} \times 100
\end{equation}
\begin{equation}
    Precision (\%)=\frac{TP}{TP+FP} \times 100
\end{equation}

Where:
\begin{itemize}
    \item True Positive (TP): The number of segments that are correctly classified the movement.
    \item True Negative (TN): The number of segments where the movement is correctly rejected.
    \item False Positive (FP): The number of segments incorrectly classified the movement, when it was         another class.
    \item False Negative (FN): The number of segments where the classifier failed to predict the correct movement. 
\end{itemize}

\begin{wrapfigure}[17]{r}{0.4\linewidth}
    \centering
  \includegraphics[width=\linewidth]{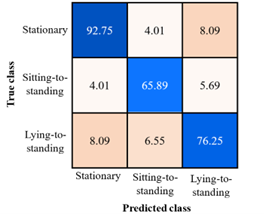}
    \caption{F1 scores for the wide ANN}
    \label{fig:enter-label}
\end{wrapfigure}

The results for these metrics for ML algorithm configurations can be found in Table \ref{fig:enter-label}. The accuracy score is a representation of the overall accuracy of the model in correctly predicting the correct movement for the 3 classes. The average validation and test accuracy values for all the ML models used was 81.4\% and 77.7\% respectively. For the recognition of the types of movements, all the algorithms had the best performance at detecting if the individual was stationary, with an average F1 score of 88.8\%. This greatly differs from the F1 scores of the sitting-to-standing and the lying-to-standing transitions of 38.7\% and 57.1\% respectively. Between algorithms, there were sizable differences between the correct recognition of movements. The standard deviations between the average values of the F1 scores for each algorithm are ±3.4\%, ±16\% and ±16.8\% for the three classes. This indicates that there is a greater spread in the recognition accuracies of for the sitting-to-standing and the lying-to-standing transitions. This can be attributed to some of the algorithms failing to detect postural movements, such as the coarse DT and coarse KNN failing to detect the sitting-to-standing movements. 

The algorithms with the highest performing accuracy scores were KNN, with an average test accuracy of 81.1\%, and ANN, with a average test accuracy of 81.6\%. These algorithms had the highest F1 scores, with ANN having a much better performance than KNN for the detection of the sitting-to-standing transition. This disparity meant that ANN had a better overall recognition performance in comparison to KNN, with an average F1 score of 6\% higher than KNN. 

The three best performing algorithms were the cubic SVM, with an average F1 score of 76.1\%, the fine KNN, with a score of 76.9\%, and the wide ANN, with a score of 78.2\%. The fine KNN algorithm had the highest performing validation and test accuracy scores, with 87.8\% and 86.4\% respectively. The quadratic SVM utilised a quadratic kernel to determine the boundaries between classes. The fine KNN was trained for the recognition of 1 neighbour. The wide ANN was a single fully connected layer with over 100 neurons between the input and output layers. 

\renewcommand{\figurename}{Table}
\setcounter{figure}{1}
\begin{figure}[!h]
    \centering
        \caption{Machine learning algorithm evaluations with accuracy scores for each model and F1 scores for each class. S = Stationary, SS = Sitting-to-standing and LS = Lying-to-standing.}
    \label{fig:enter-label}
    \vspace*{-5mm}
  \includegraphics[width=1\linewidth]{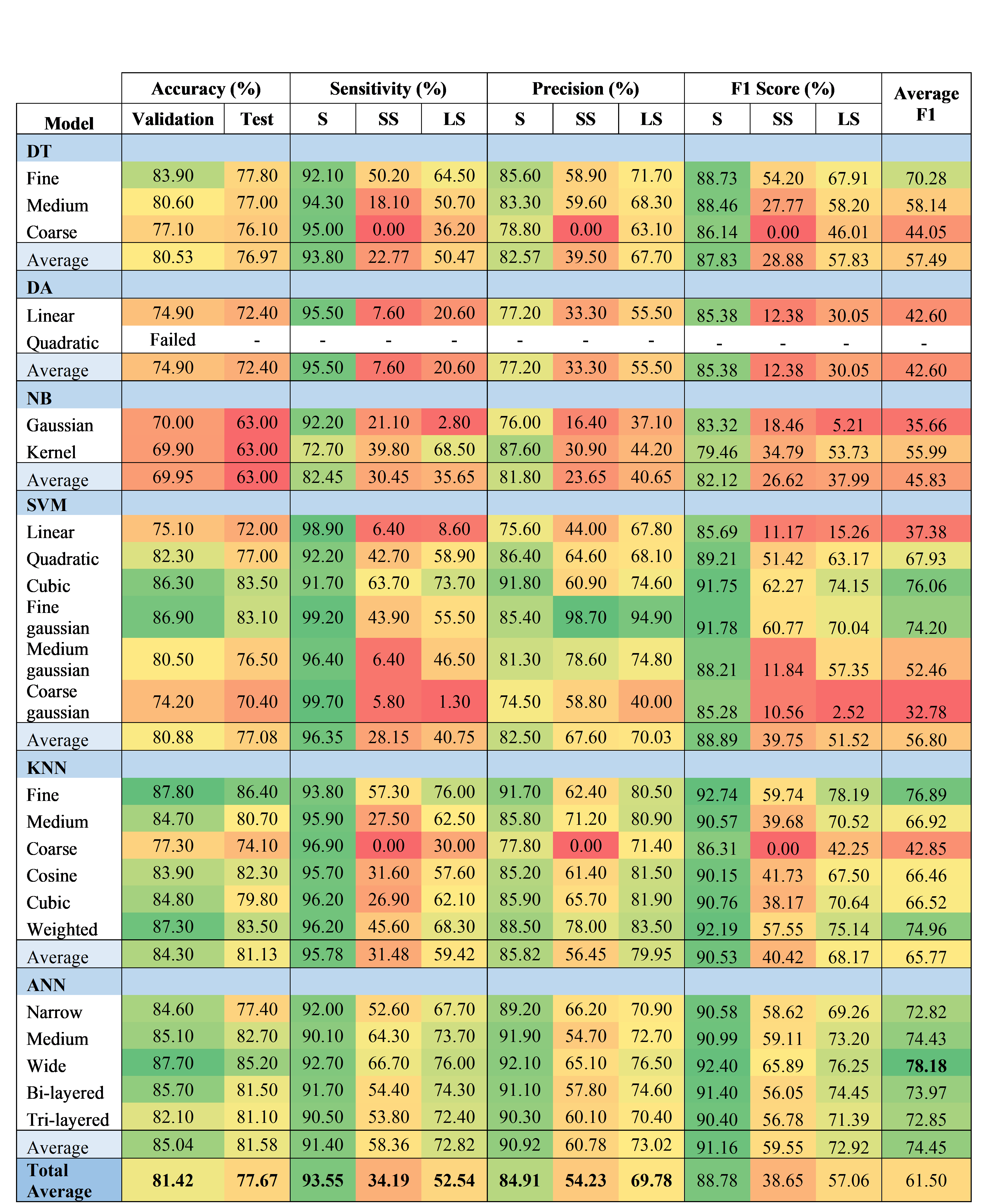}
\end{figure}

With the greatest average performance across all classes, the ML algorithm that was chosen for the recognition of movement was the wide ANN. The evaluation and test accuracies for the ANN are 87.7\% and 85.2\% respectively. Figure 19 shows the corresponding F1 scores for the correct and incorrectly movement recognition for the ANN.

\section{Discussion}
The results show that postural movement recognition is a feasible option through the analysis of PPG distortions. Between the moving and stationary measurements, the expected changes to the PPG signal occurs \cite{ref27}. As seen in Figure \ref{fig:17}, the effect of orthostatic stress and baroreflex response \cite{ref10} can be observed in the increase in the blood volume at 30 seconds due to the movement and the decrease back to the values found before the movement at 40 seconds. The differences between the magnitudes of these changes between movements, with the lying-to-standing movements having larger variations compared to the sitting-to-standing. This allowed the wide ANN to recognise the postural movements. Compared to similar HAR studies for standalone PPG signals, the proposed method had a similar performance for the recognition of the recognition of these activities \cite{ref11}. This system would allow for a general-purpose monitoring system to be used by care home staff to monitor people at risks of falls around their room. 

The greater performance for the ML algorithm in classifying the stationary PPG signals could have been due to the regular morphology of the stationary PPG pulses compared to the PPG signals distorted due to movement. For a stationary participant, the number of factors that can increase the variability of the PPG signal is less than compared to if the participant is moving. This allows for clearer patterns to be found in the features. Furthermore, the higher performance of the lying-to-standing movements compared to the sitting-to-standing can be attributed to the more complex and the increased time taken for the movement. For a younger adult,  these differences may be reduced. However, for those a risk of falls, such as elderly people, OH and age related reductions in the baroreflex \cite{ref26} increase the length and magnitudes of the distortions. The greater changes could increase the performance of the recognition of movements. However, this would have to be further investigated with elderly participants completing postural movements. 

The Chi-squared test showed that the features that had the greatest impact on recognising the types of movement were those associated with systolic point and the features with the lowest impact were those associated with the diastolic point and the dicrotic notch. This indicates that changes in the movement were more noticeable and identifiable for the features involving the systolic point. 

The overlap between class features meant that algorithms that looked for clear boundaries in large groups of data performed worse than algorithms that observed differences in classes in more complex, local groupings. Linear boundary classifiers were among the lowest performing models, with an average F1 score of 42.6\% for LDA and 37.38\% for Linear SVM. The overlap meant that ANN were able to perform particularly well, with the highest average F1 score for the three classes, as the algorithm allows for complex mapping of inputs and outputs. When comparing the wide ANN to the other single layer ANN, the larger layer size increased the overall performance of the model. The increased size allows for a greater capacity in learning more complex patterns in the dataset, thus allowing for a better ability to differentiate between classes. 

This study has several limitations. Even though the condition for the recording of postural movements were an issue that has to be considered was segment distribution imbalance. 79\% of the segments in the dataset were related to when the device wearer was stationary before the movement and after the blood volume had returned to baseline values. The remaining segments were distributed between the sitting-to-standing and lying-to-standing transition recognition sets. To give a better representation of the performance for individual classes, metric such as the precision, sensitivity and F1 score were used. However, due to the limited data set representing the movement of the individual, resampling methods that equalised the number of segments per class would greatly reduce the overall dataset training and evaluation size. This could lead to issues such as overfitting, bias and a low performance in real-world recognition of postural movements. 

Moreover, the data set was taken from young adults. To improve the accuracy of the model for the purpose of fall prevention in a care home, further work would be required to investigate the relationship between the changes in the PPG signal with movement with those more at risk of falls. 

\section{Conclusion}
This study explored the feasibility of employing a PPG sensor in combination with machine learning techniques to recognize various postural movements. A specialized hardware device was developed to capture and wirelessly transmit PPG signals from individuals as they transitioned from sitting and lying positions to standing. The data collected from these activities formed a comprehensive database used for subsequent analysis.

Machine learning algorithms DT, DA,NB,SVM,KNN and ANN was applied to classify these postural movements, with the F1 score serving as the primary metric to evaluate the effectiveness of each algorithm. Results indicated that the ANN model achieved an accuracy of 85\% and an average F1-score of 78\%, positioning it as the most effective model for this application. These findings suggest that the integration of PPG sensors with ML algorithms holds promise for accurate recognition of postural transitions in inclusive settings.


\section*{Acknowledgments}
This work was partially supported by University of Bristol internal funding. 


\end{document}